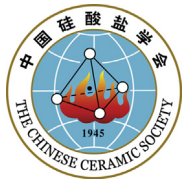

Contents lists available at ScienceDirect

# Journal of Materiomics

journal homepage: www.journals.elsevier.com/journal-of-materiomics/

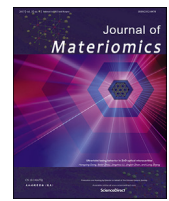

Research paper

# Crystal, ferromagnetism, and magnetoresistance with sign reversal in a EuAgP semiconductor

Qian Zhao [a], Kaitong Sun [a], Si Wu [b], Hai-Feng Li [a, *]

[a] *Institute of Applied Physics and Materials Engineering, University of Macau, Avenida da Universidade, Taipa, Macao SAR, 999078, China*
[b] *School of Physical Science and Technology, Ningbo University, Ningbo, 315211, Zhejiang, China*

**ABSTRACT**

We synthesized the ferromagnetic EuAgP semiconductor and conducted a comprehensive study of its crystalline, magnetic, heat capacity, band gap, and magnetoresistance properties. Our investigation utilized a combination of X-ray diffraction, optical, and PPMS DynaCool measurements. EuAgP adopts a hexagonal structure with the $P6_3/mmc$ space group. As the temperature decreases, it undergoes a magnetic phase transition from high-temperature paramagnetism to low-temperature ferromagnetism. We determined the ferromagnetic transition temperature to be $T_C$ = 16.45(1) K by fitting the measured magnetic susceptibility using a Curie-Weiss law. Heat capacity analysis of EuAgP considered contributions from electrons, phonons, and magnons, revealing $\eta$ = 0.03 J/(mol·K$^2$), indicative of semiconducting behavior. Additionally, we calculated a band gap of ~1.324(4) eV based on absorption spectrum measurements. The resistivity versus temperature of EuAgP measured in the absence of an applied magnetic field shows a pronounced peak around $T_C$, which diminishes rapidly with increasing applied magnetic fields, ranging from 1 to 14 T. An intriguing phenomenon emerges in the form of a distinct magnetoresistance transition, shifting from positive (*e.g.*, 1.95% at 300 K and 14 T) to negative (*e.g.*, −30.73% at 14.25 K and 14 T) as the temperature decreases. This behavior could be attributed to spin-disordered scattering.

## 1. Introduction

Magnetic semiconductors have been a subject of extensive study for decades due to their potential applications in spintronics [1]. Early examples of ferromagnetic semiconducting materials include bulk substances like europium and chromium chalcogenides [2,3]. The compound EuAgP was first synthesized in 1981 [4]. Its crystalline structure has been identified as hexagonal with the space group $P6_3/mmc$ and lattice constants $a = b$ = 439.5 pm and $c$ = 805.7 pm [4,5]. In this structure model, silver (Ag) and phosphorus (P) ions form planer honeycomb layers stacked along the $c$ direction [6]. Pöttgen and Johrendt initially proposed that EuAgP may exhibit semiconductor properties due to its electron-precise configuration, although its electrical transport characteristics had not been reported at the time [5]. More recently, based on first-principle calculations and symmetry analyses, EuAgP was projected to be a semiconductor with a band gap of ~0.18 eV in the paramagnetic state [6]. The magnetic ground state of EuAgP is ferromagnetic, featuring a metallic spin-up channel and a semiconducting spin-down channel [6]. The magnetism in EuAgP arises from Eu$^{2+}$ with spin $S$ = 7/2, orbital angular momentum $L$ = 0, and total quantum angular momentum $J$ = 7/2, where $S$, $L$, and $J$ represent the spin, orbital, and total quantum angular momentum, respectively. An ongoing issue concerns the nature of the localized $4f^7$ electrons of Eu$^{2+}$ with ferromagnetic interactions [7]. The main theoretical advancements can be traced back to Goodenough's hypothesis of cation-cation interaction through the overlap of localized 4$f$ states with 5d orbitals at neighboring Eu ions [7,8]. On the other hand, EuAgP shares similarities with ferromagnetic metals, hosting highly localized 4f electrons alongside potentially itinerant 5s electrons from Ag$^+$. According to Kataoka's research [9], significant spin fluctuations can arise in the vicinity of the Curie temperature ($T_C$) when a ferromagnetic state approaches instability, disrupting the long-period spin structure. Even in the presence of nonmagnetic scattering, critical spin fluctuations with extended wavelengths can contribute to the resistivity when the Fermi surface is small, resulting in a characteristic peak at $T_C$. The splitting of up- and down-spin bands due to magnetization can frequently turn a ferromagnetic metal with a small Fermi surface into a half-metallic material [9].

Compared to other extensively studied materials such as EuO [10], EuS [11], and EuSe [12], EuAgP has lacked precise experimental characterizations. In this study, we synthesized polycrystalline EuAgP using alumina crucibles sealed in quartz tubes. We employed room-temperature X-ray powder diffraction (XRPD) to characterize its crystalline structure and utilized a physical property measurement system (PPMS DynaCool instrument, Quantum Design) to measure magnetic properties, heat capacity, and electrical transport under applied magnetic fields. The absorption spectrum was measured across the ultraviolet to near-infrared spectral range. Our investigation centered on the analysis of the crystalline structure, magnetic properties, and the magnetoresistance (MR) effect. We found that the EuAgP compound demonstrates negative MR below $T_{NP}$, which marks the temperature at which MR transitions from negative to positive as temperature increases. This discovery offers valuable insights into the complex interplay between magnetism and electrical conductivity.

## 2. Materials and methods

### 2.1. Materials synthesis

We synthesized polycrystalline EuAgP samples by heating stoichiometric amounts (1:1:1) of europium nuggets, silver grains, and phosphorous powder in alumina crucibles. These crucibles were sealed in quartz tubes within an argon atmosphere [13]. The sealed quartz tube was initially heated gradually (90 K/h) to 1223 K and held at that temperature for 15 h. Subsequently, it was naturally cooled down to room temperature by turning off the furnace. This pre-melting step allowed the initial mixing of raw bulk metal material with the powder material. The obtained sample was ground and then annealed at 1,323 K for 15 h. After natural cooling, the sample was ball-milled for 15 min and sealed once more in a quartz tube for the third sintering at 1,323 K for 15 h, resulting in solid samples with a dark-grey color. These materials are stable in ambient air.

### 2.2. Powder diffraction

We ground the resultant material gently into a powder sample using a Vibratory Micro Mill (FRITSCH PULVERISETTE 0) for the structural study. The XRPD study was performed at room temperature with a $2\theta$ range of $10°-90°$ and a step size of $0.02°$ on an in-house X-ray diffractometer (Rigaku, SmartLab 9 kW) employing cooper $K_{\alpha 1}$ (1.54056 Å) and $K_{\alpha 2}$ (1.54439 Å) radiation at a 2:1 ratio. XRPD patterns were recorded under ambient conditions at a voltage of 45 kV and a current of 200 mA. The collected XRPD patterns were fined using the FULLPROF SUITE program [14].

### 2.3. Magnetization, specific heat, and resistivity measurements

For the measurements of DC magnetization and specific heat, we utilized a PPMS DynaCool instrument with the vibrating sample magnetometry and heat capacity options. We conducted these measurements on pressed EuAgP pellets (at 60 MPa for 15 min) within a temperature range of 1.8–350.0 K. We performed two modes of DC magnetization measurements at various applied magnetic fields, ranging from 100 Oe to 14 T: One was cooling in the absence of magnetic field (ZFC), and the other was cooling in the applied magnetic field (FC). Additionally, magnetic hysteresis loops were measured at temperatures of 1.8, 5.0, 10.0, 15.0 K and 25.0 K, with magnetic fields ranging from 0 to 14 T and back to −14 T. Specific heat measurements were carried out at magnetic fields of 0, 0.5, 1.0, 1.5, 2.0 T and 5.0 T, within a temperature range of 1.8–180.0 K. We utilized the resistivity option from Quantum Design to measure the temperature-dependent resistivity of a polycrystalline EuAgP bar (dimensions: 2.55 mm × 1.50 mm × 5.36 mm). This bar was cut from a pellet pressed at 60 MPa for 15 min. The measurements were conducted in the temperature range from 2 K to 300 K at magnetic fields ranging from 0 to 14 T using a four-probe setup. To connect the bar to the setup, SPI silver paint was used in conjunction with Au wires.

### 2.4. Optical absorption

To study the electrical band gap of the EuAgP compound, we measured its light absorption spectrum across the ultraviolet (UV) to near-infrared (NIR) spectral region, covering a wavelength range from approximately 250 nm up to 2,500 nm. This measurement was conducted using a UV–visible (Vis)–NIR spectrophotometer (model: Jasco V-770) equipped with an integrating sphere at the University of Macau in Macao SAR, China.

### 2.5. X-ray photoelectron spectroscopy

To investigate the surface chemical properties of EuAgP, we utilized high-resolution X-ray photoelectron spectroscopy (Thermo Fisher Scientific, ESCALAB 250Xi). A full-scan spectrum was acquired with an incident energy of 50.0 eV at room temperature, followed by a narrow-scan spectrum obtained at an incident energy of 20.0 eV, both with an energy step size of 0.1 eV and an etching time of 1,000 s.

## 3. Results and discussion

### 3.1. Structural study

We performed XRPD studies on EuAgP at room temperature to examine its crystal structure. The collected and refined patterns are depicted in Fig. 1a. Our data could be accurately indexed using the space group $P6_3/mmc$ within present experimental errors, which aligns with previous studies [4,13]. Given the high symmetry of the structure, there are relatively small number of refinement parameters. The goodness of fit, represented by $\chi^2$, falls within an acceptable range (see Table 1), affirming that we successfully synthesized a single-phase EuAgP compound. The structural parameters extracted from our analysis are listed in Table 1. Furthermore, Fig. 1b displays the corresponding crystalline structure within a single unit cell, where Eu atoms are situated between adjacent Ag–P layers.

### 3.2. Magnetization

Fig. 2a illustrates the relationship between magnetization and temperature. As depicted on the left axis of Fig. 2a, as the temperature decreases from 260.0 K to 1.8 K, magnetization experiences a rapid increase, starting from around 30 K, and reaching a maximum at ~15 K. Subsequently, the ZFC and FC magnetization curves diverge, with the ZFC curve decreasing quickly and the FC curve gradually rising again. This behavior signifies a weak spin freezing effect, characteristics of a magnetic phase transition associated with $Eu^{2+}$. The Curie-Weiss (CW) law is used to fit the inverse magnetic susceptibility ($\chi^{-1}$) in the pure paramagnetic (PM) state,

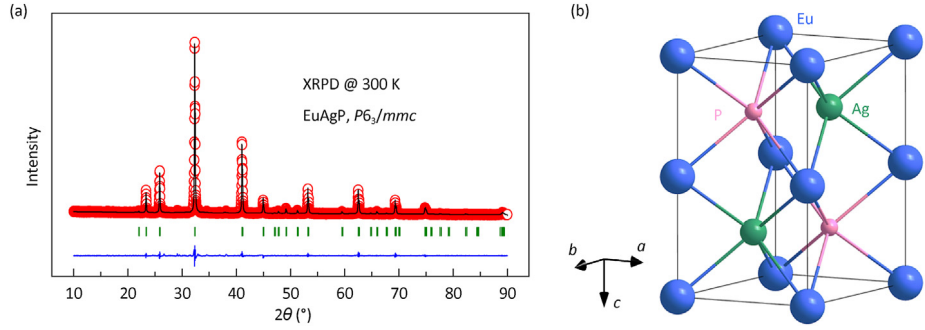

**Fig. 1.** (a) Observed X-ray powder diffraction pattern of EuAgP at 300 K is represented by circles. The solid line corresponds to the calculated pattern within the $P6_3/mmc$ space group (No. 194). Vertical bars indicate the positions of Bragg reflections, and the lower curve illustrates the difference between the observed and calculated patterns. (b) Crystal structure of EuAgP is depicted, showing a single unit cell delineated by solid lines.

**Table 1**
Refined structural parameters of EuAgP, with a hexagonal crystal structure in space group $P6_3/mmc$ (No. 194) and a unit cell containing 2 formula units ($Z = 2$), have been determined from room-temperature XRPD. These parameters include lattice constants, unit-cell volume ($V$), atomic coordinates, and goodness of fit. The Wyckoff sites for all atoms are listed. In the FULLPROF refinement process, we have retained the atomic occupation factors (OCs). The values in parentheses represent the estimated standard deviations of the (second) last significant digit. The refinement statistics are as follows: $R_p = 5.04$, $R_{wp} = 6.80$, $R_{exp} = 3.08$, and $\chi^2 = 4.88$.

| $a$ (Å)     | $b$ (Å)     | $c$ (Å)      | $V$ (Å$^3$) | $\alpha$ (=$\beta$) (°) | $\gamma$ (°) |
| ----------- | ----------- | ------------ | ----------- | ---------------------- | ------------ |
| 4.39379(8)  | 4.39379(8)  | 8.05870(17)  | 134.733(5)  | 90                     | 120          |
| Atom        | Site        | $x$          | $y$         | $z$                    | OCs          |
| Eu          | 2a          | 0.00         | 0.00        | 0.00                   | 0.0833       |
| Ag          | 2d          | 0.3333       | 0.6667      | 0.75                   | 0.0833       |
| P           | 2c          | 0.3333       | 0.6667      | 0.25                   | 0.0833       |

$$\chi^{-1}(T) = \frac{3k_B(T - \theta_{CW})}{N_A \mu_{eff}^2}, \qquad (1)$$

where $k_B = 1.38062 \times 10^{-23}$ J/K is the Boltzmann constant, $\theta_{CW}$ is the PM CW temperature, $N_A = 6.022 \times 10^{23}$ mol$^{-1}$ is the Avogadro's constant, and $\mu_{eff} = g\mu_B \sqrt{J(J+1)}$ is the effective PM moment, $\mu_B$ is the Bohr magneton. We applied a fitting to the magnetization data in the temperature range of 16.5–200.0 K using Eq. (1) and extrapolated the fitting to $M(\theta_{CW}) = 0$, as illustrated in the inset and on the right axis of Fig. 2a. The CW fitting results indicate an effective PM moment of $\mu_{eff} = 6.21$ $\mu_B$ and a PM Curie temperature $\theta_{CW} = 16.46$ K. Notably, the experimental effective PM moment is smaller than the corresponding theoretical value (7.94 $\mu_B$) for Eu$^{2+}$ (with electron configuration 4$f^7$ and quantum numbers $S = 7/2$, $L = 0$, and $J = 7/2$), as well as the reported value of 7.52 $\mu_B$ by Tomuschat et al. [4,5]. This difference may arise from Eu valence fluctuations between $Eu^{2+}$ and $Eu^{3+}$ states or a potential Eu dificiency.

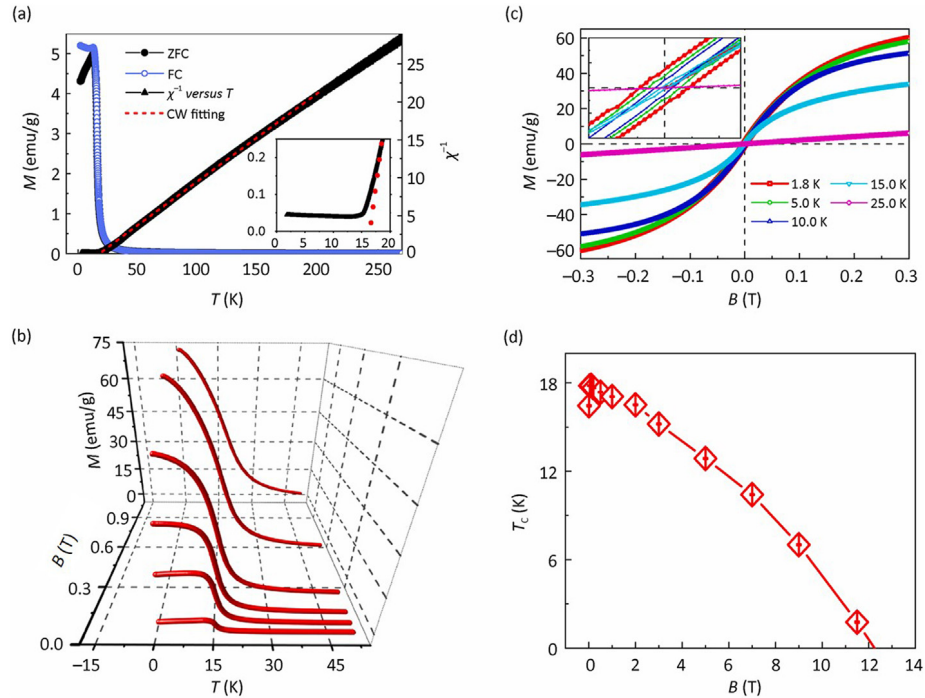

**Fig. 2.** (a) Temperature-dependent ZFC magnetization (solid circles) and FC magnetization (hollow circles) on the left vertical axis ($M$), along with the calculated ZFC inverse magnetic susceptibility (solid triangles) on the right vertical axis ($\chi^{-1}$). The measurements were taken under an applied magnetic field of 100 Oe in the temperature range of 1.8–260.0 K. The dashed line represents the fit using a Curie-Weiss law. Inset provides a view of where the Curie-Weiss curve intercepts the horizontal temperature axis. (b) Temperature dependence of ZFC magnetization under applied magnetic fields of 100, 500, 1000, 2000, 5,000 Oe and 10,000 Oe, ranging from 1.8 K to 50.0 K. (c) Magnetization as a function of magnetic field, ranging from −0.3 T to 0.3 T at temperatures of 1.8, 5.0, 10.0, 15.0 K and 25.0 K. Inset offers a closer view of the hysteresis loops. (d) Variation of the Curie temperature ($T_C$) with applied magnetic field, extracted through Curie-Weiss fitting under different applied magnetic fields.

Magnetization measurements as a function of temperature at various applied magnetic fields are shown in Fig. 2b. As the applied magnetic field increases, the observed magnetization anomaly becomes less pronounced, and the kink progressively weakens. Fig. 2c shows the ZFC magnetization as a function of applied magnetic field, ranging from −0.3 T to 0.3 T at various temperatures. The magnetization at temperatures below the ferromagnetic (FM) transition temperature ($T_C$ = 16.45(1) K) exhibits a non-linear field dependence, characteristic of ferromagnetism. In the inset of Fig. 2c, very small hysteresis loops are visible, indicating that EuAgP exhibits characteristics of a soft ferromagnet. To gain a deeper understanding of the magnetic phase transition under varying magnetic fields, we applied Eq. (1) to fit the M(T) curves and extracted the corresponding $T_C$ values. As depicted in Fig. 2d, the $T_C$ curve as a function of applied magnetic field adheres to a square law [15]. Our magnetization study provides evidence that the magnetic ground state of EuAgP is indeed ferromagnetic.

### 3.3. Heat capacity

The heat capacity measurements are shown in Fig. 3. A λ-shaped transition, as shown in Fig. 3a, takes place at $T_C$ = 16.45(1) K, which is consistent with the FM transition observed in our magnetic susceptibility measurements. To analyze this data, we applied a fitting model to the specific heat curve as a function of temperature,

$$C = C_{ele} + C_{pho} + C_{mag}, \quad (2)$$

where $C_{ele} = \eta T$ represents the contribution from electrons, $C_{pho}$ stands for the contribution from phonons, and $C_{mag}$ corresponds to the contribution from magnons [16,17]. In the calculation of $C_{pho}$,

$$\begin{aligned}C_{pho} &= C_{Debye} + C_{Einstein} \\ &= 9nR(1-d)\left(\frac{T}{\Theta_D}\right)^3 \int_0^{\frac{\Theta_D}{T}} \frac{x^4 e^x}{(e^x-1)^2} dx \\ &\quad + 3nRd\left(\frac{\Theta_E}{T}\right)^2 \frac{e^{\frac{\Theta_E}{T}}}{\left(e^{\frac{\Theta_E}{T}}-1\right)^2},\end{aligned} \quad (3)$$

where $n = 3$ represents the number of atoms in the chemical formula, $R = 8.314$ J/mol/K stands as the ideal gas constant, and $\Theta_D$ and $\Theta_E$ denote the Debye temperature and Einstein temperature, respectively. The coefficients $d$ and $(1−d)$ signify the Einstein and Debye phonon contributions, respectively. Over the temperature range of 1.8–160.0 K, the Debye-Einstein model exhibits a good agreement with the nonmagnetic portion of specific heat (refer to Fig. 3a). This alignment yields a parameter value of $\eta = 0.03$ J/(mol·$K^2$). The small $\eta$ value aligns with the semiconducting behavior of EuAgP, implying a low density of electronic states at the Fermi level [18]. By extrapolating and subtracting the electron and phonon contributions from the low-temperature data, we were able to calculate the magnetic contribution to heat capacity (see Fig. 3a). This further confirmed the magnetic behavior of EuAgP at low temperatures.

Fig. 3b depicts the adiabatic heat capacity measured at different magnetic fields, with the inset focusing on the specific variations around the FM transition temperature $T_C$. A noticeable shift in the heat capacity versus magnetic field curves was observed, indicating the nature of this transition as a magnetic phase transition. The influence of the applied magnetic field on heat capacity diminishes as the temperature drops below $T_C$, suggesting a decreasing impact of the magnetic field on the order parameter (i.e., magnetization) and specific heat capacity. Meanwhile, the observed anomaly, in the form of a λ-type kink, gradually smoothens as the applied magnetic field increases. This phenomenon may be caused by the demagnetizing effect [19]. The effective internal magnetic field ($B_{in}$) is weaker than the externally applied field $B$, and the external field becomes increasingly filtered as magnetization grows at lower temperatures. The disappearance of the λ-type kink can be attributed to the fact that the data was collected under high magnetic fields, where field penetration is nearly complete around $T_C$ [19–21].

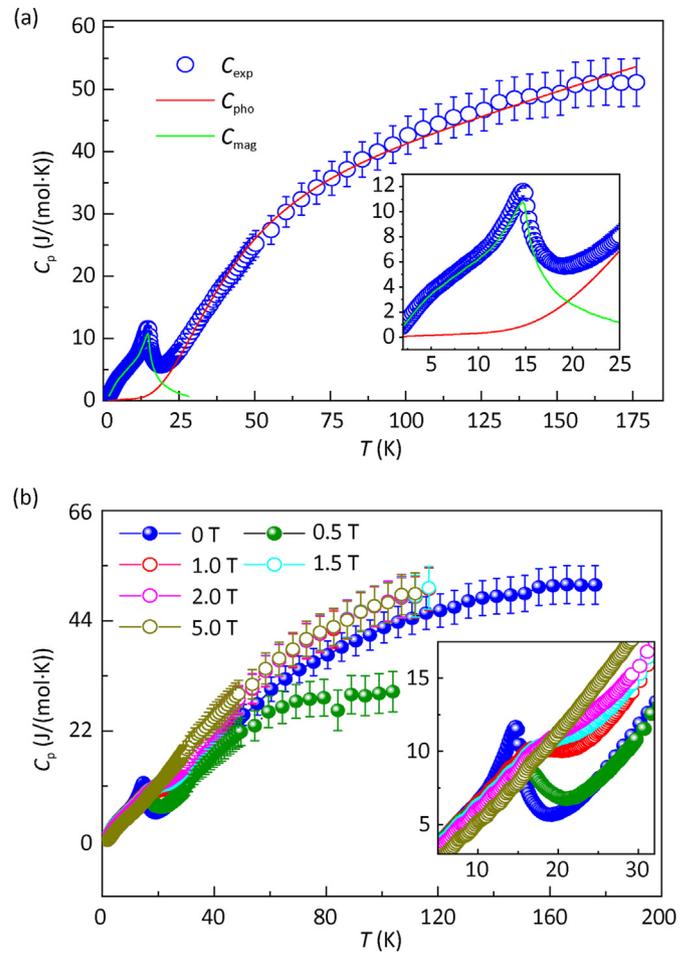

**Fig. 3.** (a) Heat capacity of EuAgP as a function of temperature at 0 T (hollow circles). The red line, covering the range from 1.9 K to 176.0 K, represents a fit that accounts for the phonon contribution only to the measured heat capacity. The magnetic contribution extracted from the data is depicted by the green line within the range of 1.9–28.0 K. (b) Temperature-dependent heat capacity for various applied magnetic fields ranging from 0 to 5 T, as indicated. Inset provides a view of how the λ-peak changes with applied magnetic fields within the temperature range of 10–30 K.

### 3.4. Resistivity

Fig. 4a shows the measured resistivity of EuAgP over a temperature range of 2–300 K with an applied magnetic field $B$ perpendicular to the direction of electric current. At lower temperatures, the resistivity exhibits a pronounced peak around $T_C$ = 16.45(1) K, in line with the concept of spin-disorder scattering [22]. This behavior aligns with observation in the traditional FM semiconductor EuO [10]. Moreover, our experimental results are consistent with carrier mobility calculations [22]: there is a minimum carrier mobility at $T_C$, leading to a maximum resistivity at $T_C$. Similar observations were reported in semiconducting materials like $CdCr_2Se_4$ [23] and $Eu_{1-x}Gd_xSe$ compounds [7,24]. As the temperature approaches $T_C$, electrons encounter a fluctuating potential caused by on-site spin exchange interactions. The presence of disordered spins introduces an additional impurity-like term to ferromagnets. When electrons, carrying most of the current, interact with an atom with a reduced spin moment, they experience significant scattering due to the higher local exchange potential [25]. Consequently, the decrease in resistivity below $T_C$ can be attributed to the reduction in spin-dependent scattering, brought about by the establishment of magnetic order. The resistivity curve exhibits a re-increase in resistivity at extremely low temperatures (<10 K) (see Fig. 4a). This phenomenon results from the emergence of the half-metallic state, in which the ↓-spin band is empty at low temperatures [9]. Simultaneously, the semiconducting ground state emerges, leading to a decrease in resistivity with increasing temperature. The peak is significantly suppressed as the applied magnetic field increases, indicating negative MR.

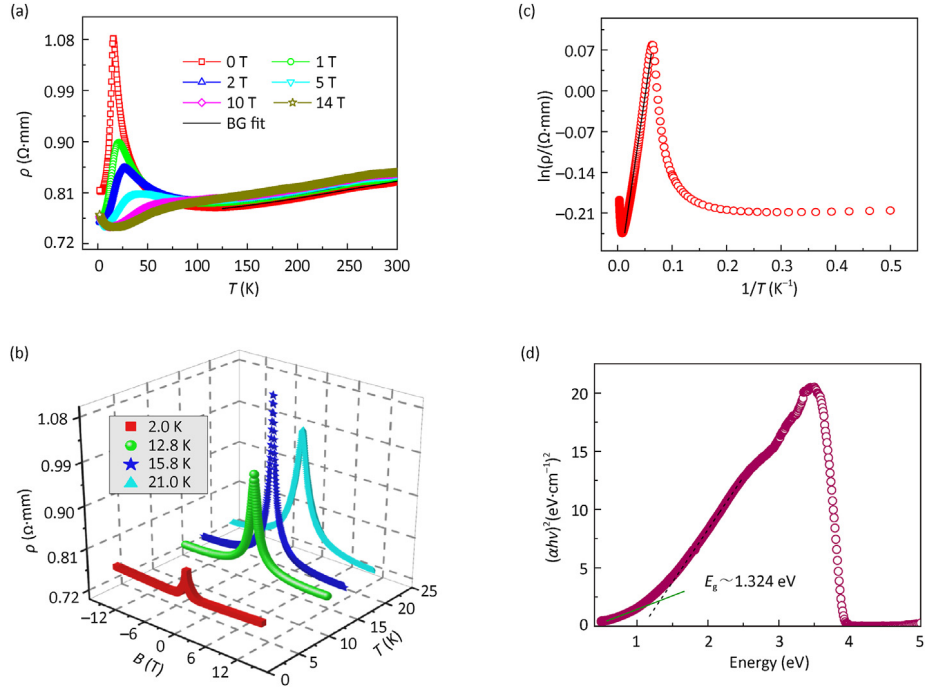

**Fig. 4.** (a) Measured resistivity of EuAgP under various applied magnetic fields, denoted as B (0, 1, 2, 5, 10 T and 14 T), with the magnetic field oriented perpendicular to the direction of electric current. The temperature range covered is from 2 K to 300 K. The black line represents a fitting model described in the text, known as the BG fit. (b) Field dependency of electrical resistivity, ranging from −14 T to 14 T, at temperatures of 2.0, 12.8, 15.8 K and 21.0 K. (c) Temperature-dependent resistivity in the absence of a magnetic field, accompanied by a fitting model to Eq. (4) (solid line), as explained in the text. (d) Tauc plot of the EuAgP compound. The dashed line represents the linear fit, while the fitted solid olive line serves as the baseline for determining the band gap energy value (for further details, please refer to the main text).

The resistivity of EuAgP at temperatures of 2.0, 12.8, 15.8 K and 21.0 K as a function of the applied magnetic field ranging from −14 T to 14 T is displayed in Fig. 4b. These temperatures are below (2.0 K and 12.8 K), around (15.8 K), and above (21.0 K) the FM transition temperature $T_C = 16.45(1)$ K. Consistent with Fig. 4a, external magnetic fields readily suppress the resistivity anomaly resulting from critical spin fluctuations around $T_C$, leading to negative MR. Theoretically, the change in resistance under an applied magnetic field occurs due to the reduction of the mean free path in the current direction, with electrons completing a significant fraction of a cyclotron orbital before being scattered [25].

For semiconductors,

$$\rho \propto \exp\left(\frac{E_g}{2k_BT}\right), \quad (4)$$

where $E_g$ could be interpreted as the band gap energy [26]. Fig. 4c illustrates a plot of $\ln\rho$ against $T^{-1}$ for EuAgP. Assuming that all carriers are predominantly thermally excited across the band gap, both electron and hole concentrations become essentially equivalent in the low-temperature $1/T$ (high-temperature $T$) region [26]. Consequently, we carried out a linear fit to the semiconductor behavior in the critical region just above the FM transition point, focusing on the high-temperature (low $1/T$) part of the $\ln\rho$ versus $T^{-1}$ curve using Eq. (4) (as depicted in Fig. 4c). This analysis reveals an estimated band gap energy of $E_g = 1.087(8)$ eV.

### 3.5. Applied field dependent magnetoresistance

We calculated the MR and presented it as a function of the applied magnetic field in Fig. 5a using the following equation,

$$MR(B) = \frac{\rho(B) - \rho(B=0)}{\rho(B=0)} \times 100\%, \quad (5)$$

At a temperature of 15.8 K, a small negative MR (∼−30.54%) is obtained. The MR effect is most prominent in the vicinity of $T_C$. However, at 200 K, the resistivity does not respond to the applied magnetic field. In the PM phase, the magnetization-dependent behaviors of the resistivity at different temperatures remain largely consistent. This behavior is indicative of resistivity influenced by spin fluctuations [9]. As discussed earlier, a magnetic field can alter the degree of spin disorder in magnetic semiconductors, thereby affecting resistivity. Additionally, changes in carrier concentration due to the applied magnetic field may also contribute to MR [22]. It is believed that in materials with significant scattering, MR is minimal (e.g., ≈1% at 1 T), whereas in semimetals and semiconductors with rapid electron mobility, MR can be substantially larger [25]. The behavior of EuAgP does not conform to Kohler's rule,

$$\frac{\rho(B)}{\rho(0)} = f\left(\frac{B}{\rho(0)}\right)^2, \quad (6)$$

which describes the field dependence of a metal's resistivity [25]. This inconsistence rules out the possibility of EuAgP being a metal.

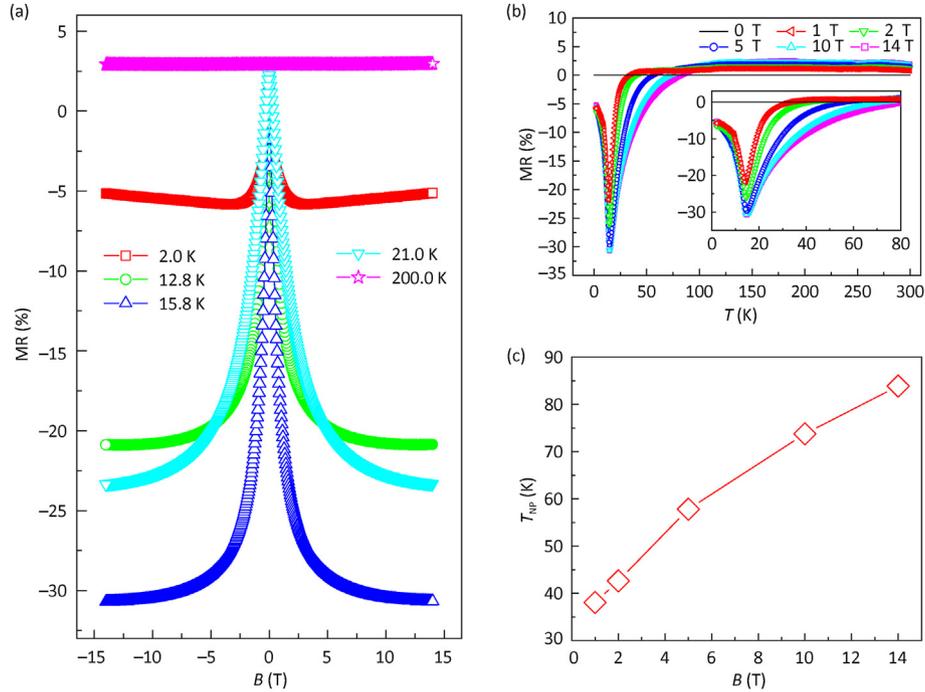

**Fig. 5.** (a) Field dependence of the calculated magnetoresistance percentage, MR%, for EuAgP at various temperatures, as indicated. (a) Temperature-dependent MR effect of EuAgP under various applied magnetic fields, including 0, 1, 2, 5, 10 T and 14 T. Inset focuses on a narrower temperature range (0–80 K) around $T_C$. (b) $T_{NP}$ as a function of the magnetic field B. Here, $T_{NP}$ indicates the temperature at which the MR value transitions from positive to negative as the material is cooled.

### 3.6. Temperature dependent magnetoresistance

Fig. 5b shows the calculated temperature-dependent MR effect of EuAgP under applied magnetic fields. As the temperature decreases, the MR maintains a positive value above $T_{NP}$ (temperature point of MR transition), after which it transitions into a negative value below $T_{NP}$. Below $T_{NP}$, the MR value progressively decreases with increasing applied magnetic field, for example, MR values of −21.87% (at 14.25 K and 1 T), −29.69% (at 14.25 K and 5 T), and −30.73% (at 14.25 K and 14 T). Conversely, above $T_{NP}$, the MR value increases as the applied magnetic field strength grows. For instance, MR values of 0.78% (at 300 K and 1 T), 1.51% (at 300 K and 5 T), and 1.95% (at 300 K and 14 T). Notably, it is observed that the value of $T_{NP}$ rises with increasing applied magnetic field, as shown in Fig. 5c. Colossal negative MR effects have been reported in a $La_{1.37}Sr_{1.63}Mn_2O_7$ single crystal, where the MR varies with both increasing applied magnetic field and changing temperature [27]. In the case of the GdSi compound, the negative MR value transitions to positive around the antiferromagnetic transition temperature, attributed to the formation of magnetic polarons [28]. In the context of EuAgP, we have observed distinctive temperature- and magnetic-field-dependent behaviors.

### 3.7. Determination of the band gap energy

The band gap energy is a crucial parameter in assessing the photo-physical properties of a semiconductor. To experimentally determine the band gap energy of the EuAgP compound, we employed the Tauc-plot method [29], i.e.,

$$(\alpha \bullet h\nu)^{\frac{1}{n}} = C(h\nu - E_g), \quad (7)$$

where $\alpha$ represents the energy-dependent absorption coefficient, $h = 6.626 \times 10^{-34}$ J·s is the Planck constant, $\nu$ is the photon's frequency, $n = 1/2$ for the direct transition band gap, C is a constant, and $E_g$ denotes the band gap energy. The Tauc plot for EuAgP is presented in Fig. 4d, revealing a steep increase in absorption with rising energy below ~3.5 eV, a characteristic feature of semiconductors. We fit the linear portions of the data using Eq. (7), from which we derived an estimate for the band gap width through a dual crossed-line fitting approach, i.e., intersection value of the two linear fitting lines, denoted as $E_g = 1.324(4)$ eV. This result is roughly consistent with the band gap value (1.087 eV) obtained from our resistivity study (Fig. 4c).

### 3.8. Europium valence fluctuation

Chemical shifts or alterations in binding energy observed in X-ray photoelectron spectroscopy spectra are indicative of different oxidation states of europium. Fig. 6 displays the Eu 3d X-ray photoelectron spectroscopy of the EuAgP surface. In accordance with the X-ray photoelectron spectroscopy handbook [30], a rigorous quantitative analysis of the spectrum identifies peaks at 1125.0(9) eV and 1154.7(6) eV, corresponding to $Eu^{2+}$ $3d_{5/2}$ and $3d_{3/2}$ states, respectively, while those at 1134.6(1) eV and 1164.5(1) eV relate to $Eu^{3+}$ $3d_{5/2}$ and $3d_{3/2}$ states, respectively. At the current resolution, the extracted content ratio between $Eu^{3+}$ and $Eu^{2+}$ is approximately 9:1.

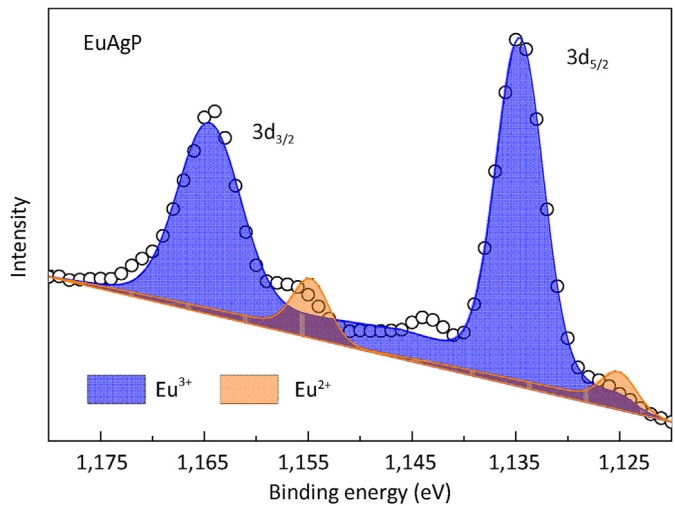

Fig. 6. X-ray photoelectron spectroscopy spectrum (circles) captures the Eu 3d-edge characteristics at the surface of EuAgP. The spectrum reveals discernible features: including the $3d_{3/2}$ and $3d_{5/2}$ peaks corresponding to $Eu^{2+}$ (filled orange area) and $Eu^{3+}$ (filled blue area).

$Eu^{3+}$ quantum numbers ($4f^6$, $^7F_0$) are $S = 3$, $L = 3$, and $J = 0$, resulting in theoretical saturation moment ($M_{sat-theo}^{Eu^{3+}}$) and effective paramagnetic moment ($\mu_{eff-theo}^{Eu^{3+}}$) values of zero. Conversely, for $Eu^{2+}$ ($4f^7$, $^8S_{7/2}$), $S = 7/2$, $L = 0$, and $J = 7/2$, leading to $M_{sat-theo}^{Eu^{2+}} = 7\,\mu_B$ and $\mu_{eff-theo}^{Eu^{2+}} = 7.937\mu_B$ [31,32]. Thus, the primary magnetic contribution stems from $Eu^{2+}$. Consistent with preceding discussions, magnetic data supports the prevalence of $Eu^{2+}$ at low temperatures. Consequently, a significant proportion of $Eu^{2+}$ is inferred at low temperatures, contrasting with the dominate presence of $Eu^{3+}$ at room temperature. This suggests the likelihood of Eu valence fluctuation, with a transformation from $Eu^{3+}$ to $Eu^{2+}$ as temperature decreases. A more comprehensive study would entail X-ray photoelectron spectroscopy measurements conducted at low temperatures, a capability presently beyond our experimental conditions.

## 4. Conclusions

In summary, we conducted a comprehensive analysis of EuAgP through XRPD, magnetic, specific heat, optical, and electrical resistance measurements. The synthesized EuAgP polycrystal belongs to a hexagonal structure with the space group $P6_3/mmc$. We refined the collected XRPD pattern to extract crystallographic information for EuAgP. Our investigations confirmed that EuAgP is a ferromagnet with Curie temperature ($T_C$) of 16.45(1) K and effective magnetic moment ($\mu_{eff}$) of 6.21 $\mu_B$. Hysteresis loop measurements demonstrate a weak FM behavior. The correlation between $T_C$ and applied magnetic field follows the square law characteristic of ferromagnetic materials. Through specific heat analysis that considered contributions from electrons, phonons, and magnons, we determined $\eta = 0.03$ J/(mol·K$^2$), indicative of its semiconductor nature. We observed that the application of a magnetic field had a suppressing effect on the specific heat due to demagnetizing effect. The temperature-dependent resistivity of EuAgP exhibits a prominent peak at $T_C$, which was significantly reduced when subjected to applied magnetic fields, a phenomenon attributed to spin-disorder scattering and resulting in a negative MR. Using the Tauc-plot method, we estimated the band gap energy ($E_g$) to be 1.324(4) eV, which approximately aligns with the findings from our resistivity study.

## Credit author statement

Q.Z., K.T.S., and S.W. synthesized the materials and carried out the measurements. All authors discussed and analyzed the results. Q.Z. and H.-F.L. performed figure development and wrote the main manuscript text. All authors reviewed the paper. H.-F.L. conceived and directed the project.

## Declaration of competing interest

The authors declare that they have no known competing financial interests or personal relationships that could have appeared to influence the work reported in this paper.

## Acknowledgments

This work was supported by the Science and Technology Development Fund, Macao SAR (File Nos. 0090/2021/A2 and 0049/2021/AGJ), University of Macau (MYRG2020-00278-IAPME), and the Guangdong-Hong Kong-Macao Joint Laboratory for Neutron Scattering Science and Technology (Grant No. 2019B121205003).

## References

[1] Žutić I, Fabian J, Das Sarma S. Spintronics: fundamentals and applications. Rev Mod Phys 2004;76:323–410.
[2] Dietl T. Ferromagnetic semiconductors. Semicond Sci Technol 2002;17: 377–92.
[3] Kong T, Stolze K, Ni D, Kushwaha SK, Cava RJ. Anisotropic magnetic properties of the ferromagnetic semiconductor CrSbSe$_3$. Phys Rev Mater 2018;2:014410.
[4] Tomuschat C, Schuster H. ABX-Verbindungen mit modifizierter Ni$_2$In-Struktur/ABX-Compounds with a Modified Ni$_2$In Structure. Z Naturforsch B Chem Sci 1981;36:1193–4.
[5] Pöttgen R, Johrendt D. Equiatomic intermetallic europium compounds: syntheses, crystal chemistry, chemical bonding, and physical properties. Chem Mater 2000;12:875–97.
[6] Ge Y, Jin Y, Zhu Z. Ferromagnetic weyl metal in EuAgP. Materials Today Physics 2022;22:100570.
[7] S Von Molnar S Methfessel. Giant negative magnetoresistance in ferromagnetic Eu$_{1-x}$Gd$_x$Se. J Appl Phys 1967;38:959–64.
[8] Goodenough J. Magnetism and the chemical bond, inorganic chemistry section. Interscience Publishers; 1963.
[9] Kataoka M. Resistivity and magnetoresistance of ferromagnetic metals with localized spins. Phys Rev B 2001;63:134435.
[10] Penney T, Shafer MW, Torrance JB. Insulator-Metal transition and Long-Range magnetic order in EuO. Phys Rev B 1972;5:3669–74.
[11] Wachter P. Chapter 19 europium chalcogenides: EuO, EuS, EuSe and EuTe. In: Alloys and intermetallics, volume 2 of *Handbook on the Physics and Chemistry of rare earths*. Elsevier; 1979. p. 507–74.
[12] Moodera JS, Meservey R, Hao X. Variation of the electron-spin polarization in EuSe tunnel junctions from zero to near 100% in a magnetic field. Phys Rev Lett 1993;70:853–6.
[13] Johrendt D, Felser C, Huhnt C, Michels G, Schäfer W, Mewis A. Tuning the valence in ternary Eu-pnictides: the series EuPd$_{1-x}$Ag$_x$P and EuPd$_{1-x}$Au$_x$As. J Alloys Compd 1997;246:21–6.
[14] Rodríguez-Carvajal J. Recent advances in magnetic structure determination by neutron powder diffraction. Phys B Condens Matter 1993;192:55–69.
[15] Arrott A. Existence of a critical line in ferromagnetic to paramagnetic transitions. Phys Rev Lett 1968;20:1029.
[16] Tari A. The specific heat of matter at low temperatures. World Scientific; 2003.
[17] Bryan J Daniel, Trill H, Birkedal H, Christensen M, Srdanov VI, Eckert H, et al. Magnetic phase diagram of Eu$_4$Ga$_8$Ge$_{16}$ by magnetic susceptibility, heat capacity, and mössbauer measurements. Phys Rev B 2003;68:174429.
[18] Hooda M, Pavlosiuk O, Hossain Z, Kaczorowski D. Magnetotransport properties of the topological semimetal SrAgBi. Phys Rev B 2022;106:045107.
[19] Griffiths RB. Ferromagnetic heat capacity in an external magnetic field near the critical point. Phys Rev 1969;188:942–7.

[20] Simons DS, Salamon MB. Specific heat and resistivity of gadolinium near the curie point in external magnetic fields. Phys Rev B 1974;10:4680–6.
[21] Glorieux C, Thoen J, Bednarz G, White MA, Geldart DJW. Photoacoustic investigation of the temperature and magnetic-field dependence of the specific-heat capacity and thermal conductivity near the curie point of gadolinium. Phys Rev B 1995;52:12770–8.
[22] Haas C. Spin-Disorder scattering and magnetoresistance of magnetic semiconductors. Phys Rev 1968;168:531–8.
[23] Lehmann H, Harbeke G. Semiconducting and optical properties of ferromagnetic $CdCr_2S_4$ and $CdCr_2Se_4$. J Appl Phys 1967;38:946.
[24] Methfessel S. Exchange interactions of the conduction electrons in lanthanide semiconductors. Z Angew Phys 1965;18:414–32.
[25] Coey JM. Magnetism and magnetic materials. Cambridge university press; 2010.
[26] Garcia N, Damask A, Schwarz S. Semiconductors. New York, NY: Springer New York; 1998. p. 437–63.
[27] Wu S, Zhu Y, Xia J, Zhou P, Ni H, Li H-F. Colossal negative magnetoresistance effect in a $La_{1.37}Sr_{1.63}Mn_2O_7$ single crystal grown by Laser-Diode-Heated Floating-Zone technique. Crystals 2020;10:547.
[28] Li H, Xiao Y, Schmitz B, Persson J, Schmidt W, Meuffels P, et al. Possible magnetic-polaron-switched positive and negative magnetoresistance in the GdSi single crystals. Sci Rep 2012;2:750.
[29] Makuła P, Pacia M, Macyk W. How to correctly determine the band gap energy of modified semiconductor photocatalysts based on UV–Vis spectra. J Phys Chem Lett 2018;9:6814–7.
[30] Wagner CD. Handbook of X-ray photoelectron spectroscopy: a reference book of standard data for use in X-ray photoelectron spectroscopy, volume Physical Electronics Division. Eden Prairie, MN: Perkin-Elmer Corporation; 1979.
[31] Zhu Y, Xia J, Wu S, Sun K, Yang Y, Zhao Y, et al. Crystal growth engineering and origin of the weak ferromagnetism in antiferromagnetic matrix of orthochromates from te orbital hybridization. iScience 2022;25:104111.
[32] Zhu Y, Zhou P, Sun K, Li H-F. Structural evolution of single-crystal $RECrO_3$ (RE = Y, Eu–Lu) orthochromates. J Solid State Chem 2022;313:123298.

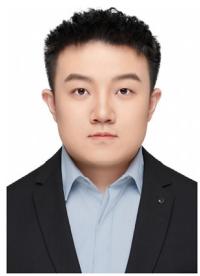

**Qian Zhao** received his B.E. degree from South China University of Technology in 2020. He is currently a Ph.D. student at the Institute of Applied Physics and Materials Engineering, University of Macau. His research focuses on the synthesis and characterization of innovative magnetic materials.

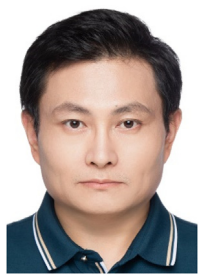

**Hai-Feng Li** received his Ph.D. degree in condensed matter science from the Research Center Juelich and RWTH Aachen University, Germany, in 2008. He was honored with an "Auszeichnung (Distinction)" for his Ph.D. thesis and received the "Borchersplakette 2009" at RWTH. Following his Ph.D., he held postdoctoral positions at the Max Planck Institute for Solid State Research (Germany, 2007–2008) and the Ames Laboratory (USA, 2008–2010). Since 2011, he has been affiliated with RWTH and the Juelich Center for Neutron Science (JCNS-2), during which he served as a long-term visitor to the Institute Laue-Langevin (France, 2011–2014). Before joining the University of Macau, he held the position of Marie Curie-COFUND CONEX Professor at UC3M, Spain. His research activities primarily focus on synthesizing advanced materials with potential macroscopic functionalities and investigating their properties through comprehensive in-house characterizations and modern neutron and X-ray scattering techniques at prominent neutron and synchrotron facilities worldwide.